\documentclass[12pt,a4paper]{article}
\usepackage{geometry}
\usepackage{setspace}
\usepackage{booktabs}  
\usepackage{graphicx}
\usepackage{longtable}
\usepackage{caption}
\usepackage{float}
\usepackage{authblk}
\usepackage{hyperref}
\usepackage[numbers]{natbib}

\title{Cultural Awareness, Stereotypes and Communication Skills in Intercultural Communication: The Algerian Participants’ Perspective}

\author{MOHAMED AMINE KADA ZAIR\thanks{CONTACT M. K. h. Email: mohamed\_zair7@hotmail.fr}}
\affil{Statistical and Stochastic Process Laboratory, Djillali Liabes University, Sidi Bel Abbes, Algeria}

\begin{document}
\maketitle
\begin{abstract}
This study explores the relationship between cultural awareness, stereotypes, and communication skills among Algerian participants working or studying in multicultural environments. A quantitative questionnaire was administered to 40 respondents to evaluate their levels of cultural awareness, the presence of stereotypical thinking, and the effectiveness of their intercultural communication skills. Results revealed that while cultural awareness was generally high, certain stereotypes still influenced the perception of others and impacted communication efficiency. Participants with higher cultural awareness demonstrated better communication skills and lower levels of stereotyping. These findings underline the importance of intercultural competence and education programs in reducing prejudice and fostering mutual understanding in diverse contexts.
\end{abstract}


\newpage

\section{Introduction}

Although economic globalisation has intensified international exchanges, cultural distinctions remain a central force shaping human interactions within professional environments. Numerous studies indicate that global integration has not erased cultural boundaries; instead, it has contributed in several contexts to the persistence and even reinforcement of cultural identities across national, regional, and social groups (\citet{lillis2010}; \citet{featherstone1995}). Recent scholarship confirms that cultural differentiation continues to play an essential role in shaping behaviour and communicative expectations within global workplaces, even in digitally mediated environments (\citet{yousef2024}; \citet{penaacuna2025}). These enduring cultural patterns continue to influence how individuals perceive messages, interpret behaviour, and engage in communicative exchanges inside multinational corporations (\citet{ferraro2002}).

Intercultural communication research consistently highlights that communicative practices are deeply embedded in culturally informed cognitive frameworks. Foundational contributions such as Hall’s distinction between high- and low-context communication (\citet{hall1976}) and Hofstede’s cultural dimensions (\citet{hofstede2001}) demonstrate how values, expectations, and interactional norms differ across societies. Building on this foundation, contemporary studies emphasise the relevance of cultural awareness not only in interpersonal communication but also in technologically mediated contexts, including AI-supported interaction systems where cultural bias may be amplified (\citet{kannen2024}). Scholars across intercultural communication and management also stress that cultural awareness—the understanding of one’s own cultural assumptions and those of others—is a necessary competence for effective communication in international business (\citet{deardorff2006} ; \citet{spenceroatey2009}). This awareness allows individuals to recognize the implicit cultural filters shaping their interpretations and responses.

At the same time, research has shown that communication can be significantly affected by stereotypes, which function as simplified cognitive representations of cultural groups. While stereotypes may help individuals navigate unfamiliar social environments, they often distort perception and create barriers to accurate understanding (\citet{gudykunst2003} ; \citet{mcgarty2002}). Recent empirical evidence further suggests that stereotype activation in global teams can be intensified by virtual work dynamics and cross-border collaboration challenges (\citet{anjum2024}. In workplace settings, stereotype-based expectations can lead to misjudgments about colleagues’ intentions, competencies, or communication styles, thereby undermining collaboration and organizational cohesion (\citet{thomas2009}).

Within multicultural and multilingual organizational contexts, these factors—cultural frameworks, cultural awareness, and stereotype-driven interpretations—interact in shaping communicative behaviour. Large-scale studies in cross-cultural management, such as the GLOBE project (\citet{house2004}), have further demonstrated how variations in cultural cognition influence leadership perception, conflict resolution, and message interpretation. More recent investigations continue to show that global team efficiency depends heavily on the ability to navigate cultural diversity through advanced communication skills and self-reflexive intercultural competence (\citet{dalib2023}; \citet{yousef2024}). As global corporations increasingly rely on diverse teams, understanding the role of culture-related variables is essential for improving communication, reducing intercultural misunderstandings, and enhancing overall organizational effectiveness (\citet{adler2008}; \citet{tingtoomey1999}; \citet{gudykunst2003}). Such scholarly insight provides a vital foundation for the development of more adaptive communication strategies suited to culturally diverse professional environments.\\
In Algeria, where people are frequently exposed to intercultural influences through media, migration, and international cooperation, understanding how cultural awareness and stereotypes affect communication skills has become essential. Previous studies have shown that stereotypes can create barriers in intercultural communication, while cultural awareness and sensitivity tend to enhance mutual understanding and cooperation.

The present research investigates how these three dimensions—cultural awareness, stereotypes, and communication skills—interact and influence one another in the Algerian context. By analyzing these relationships, the study aims to provide insight into the psychological and social mechanisms that underlie effective communication in multicultural environments.

\section{Research Aim and Theoretical Perspectives}
The main aim of the present study is to examine the extent to which culture—understood from the anthropological perspective as the body of knowledge, beliefs, moral rules, customs, and learned habits shared by members of a social group (\citet{benedict1934}; \citet{tylor1958}; \citet{hill2005}), and from the cognitive perspective as the collective programming of the mind, sets of cognitive patterns, emotional orientations, and interaction styles transmitted through symbolic systems across generations (\citet{kluckhohn1951}; \citet{geertz1973}; \citet{dandrade1984}; \citet{triandis1994}; \citet{hofstede2001}; \citet{matsumoto2006}) —shapes the perceptions and practices of communication among Algerians living abroad in multicultural environments. More specifically, the study investigates how cultural awareness, stereotypes, and communication skills interact to influence the way Algerian migrants—both workers and students—interpret, evaluate, and engage in communicative exchanges in their host countries.

The research is structured around three interrelated analytical dimensions essential for understanding intercultural communication within diasporic contexts.
The first dimension, cultural awareness, concerns the degree to which Algerian migrants recognize cultural differences, decode culturally embedded meanings, and understand that emotions, expectations, and social norms are conveyed and negotiated through communication. This dimension highlights how migrants balance adaptation to new cultural norms with the preservation of culturally shaped communicative habits.

The second dimension, stereotypes, focuses on the cultural and communicative barriers experienced by Algerians abroad, especially those arising from stereotype-based perceptions. These stereotypes may be directed toward Algerians by members of the host society or held by Algerians about other cultural groups. Examining migrants’ encounters with stereotype-driven behavior provides insight into how perceived cultural distance and pre-existing assumptions influence their understanding of intercultural interactions.

The third dimension, communication skills, relates to the strategies and competences Algerian migrants employ when navigating multicultural settings. This includes their ability to adjust speech styles, manage misunderstandings, interpret nonverbal cues, and adopt effective communicative behaviors in diverse academic, professional, and social environments. Assessing migrants’ communication skills allows for a clearer understanding of how individuals negotiate identity, maintain social relationships, and achieve communicative effectiveness across cultures.

By investigating these three complementary dimensions, the study contributes to a deeper understanding of the psychological, social, and cultural mechanisms that shape communication among Algerian migrants. It also provides empirical insight into how cultural awareness, stereotypes, and communication skills jointly influence migrants’ capacity to interact successfully within multicultural environments, thereby enriching the broader field of intercultural communication research.

\subsection{Cultural Awareness}

Cultural awareness constitutes a foundational pillar of intercultural competence and refers to the capacity of individuals to recognize, reflect upon, and understand both their own cultural frameworks and those of others (\citet{byram1997}; \citet{deardorff2006}). It involves an ongoing process of identifying culturally shaped beliefs, values, communication norms, and patterns of behavior that influence how people interpret social reality. From an anthropological perspective, cultural awareness reflects the ability to acknowledge that each cultural system provides its members with particular ways of perceiving, categorizing, and interacting with the world (\citet{geertz1973}; \citet{keesing1981}). This implies recognizing that interpretations of messages, social expectations, and relational practices are embedded in historically transmitted symbolic structures, myths, and collective experiences shared by specific groups.

From a cognitive and psychological viewpoint, cultural awareness encompasses the metacognitive understanding that perception, emotion, and communication are culturally mediated, and that individuals interpret behaviors through culturally learned schemas (\citet{triandis1994}; \citet{matsumotojuang2013}). It allows individuals to recognize the relativity of their own cultural assumptions—what \citet{bhabha1994} describes as the “de-centering of the self”—and promotes openness, empathy, and intellectual flexibility when interacting with others. Through this reflective process, individuals come to understand that cultural differences are neither deficits nor obstacles but distinct systems of meaning that shape interpersonal interaction.

In intercultural communication, cultural awareness enables individuals to navigate communicative encounters more effectively by adjusting interpretations, anticipating misunderstandings, and responding appropriately to culturally grounded expectations (\citet{gudykunst2004}; \citet{tingtoomeychung2012}). It encourages recognition of subtle differences in conversational styles, power distance relations, forms of politeness, and nonverbal cues that vary across cultural groups. Such awareness is particularly relevant in diasporic contexts where migrants must negotiate between inherited cultural identities and new cultural environments, engaging in what \citet{berry1997} describes as processes of acculturation and integration.

For Algerian migrants, cultural awareness also involves sensitivity to ultra-national diversity shaped by Arab, Berber, Mediterranean, and Islamic influences, which create multilayered cultural identities. Thus, understanding cultural awareness in this context provides insights into how individuals interpret their host cultures while simultaneously drawing upon their own cultural repertoires to engage in social, academic, and professional communication abroad. In sum, cultural awareness is not merely knowledge about cultural differences but a dynamic competence that facilitates successful interaction, reduces ethnocentric bias, and fosters meaningful participation in multicultural environments (\citet{spitzbergchangnon2009}).

\subsection{Stereotypes}

Stereotypes represent one of the most widely examined constructs in the fields of social psychology, intercultural communication, and cognitive sociology. They are generally understood as socially shared, collectively transmitted beliefs about the characteristics, behaviors, or dispositions of members of particular groups (\citet{mccrae2012}; \citet{stangor1996}). Early conceptualizations—starting with Lippmann’s foundational description of stereotypes as “pictures in our heads’’ (\citet{lippmann1922})— highlighted their role as mental shortcuts that allow individuals to reduce the overwhelming complexity of social life by filtering, structuring, and simplifying information.

From a cognitive perspective, stereotypes function as schema-based categorization mechanisms that assist individuals in organizing social information and predicting human behavior (\citet{hamilton1994}; \citet{fisketaylor2013}). They help people navigate unfamiliar situations by providing readily available expectations about others. Although such cognitive economy supports efficient processing, it frequently results in oversimplifications, distortions, and essentialist assumptions that obscure individual variability (\citet{hilton1996}). Hence, stereotypes are simultaneously functional and problematic: they enhance mental efficiency, yet they also limit accuracy and openness in social perception.

From a cultural and anthropological standpoint, stereotypes are products of historically accumulated narratives, symbolic systems, and intergroup relations (\citet{barth1969}; \citet{berger1966}). They are influenced by collective memory, political discourse, educational systems, and media representations that shape the ways social groups are imagined and evaluated (\citet{hall1997}; \citet{pickering2001}). As such, stereotype formation is not merely a cognitive habit but also a socio-cultural process through which societies define boundaries between “us’’ and “them’’ (\citet{tajfel1981}; \citet{jenkins2008}). Variables such as ethnicity, religion, nationality, gender, and social class often serve as the primary axes along which stereotypical distinctions emerge and become normalized.

In intercultural communication, stereotypes play a decisive role because they frame expectations, constrain interpretations, and influence communicative behavior. They may operate implicitly—shaping judgments without conscious awareness—or explicitly, guiding overt attitudes and behaviors toward members of other cultural groups (\citet{devine1989}; \citet{matsumotojuang2013}). Negative stereotypes can lead to prejudice, discriminatory behavior, and communication breakdowns, while positive or neutral stereotypes may still result in misread cues, inaccurate assumptions, and superficial interactions. Consequently, understanding stereotypes is essential for analyzing how individuals from culturally diverse backgrounds construct meaning, attribute intentions, and manage interactions in multicultural settings.

Exploring stereotypes among Algerian migrants, in particular, offers valuable insight into the cultural and psychological foundations of intergroup perception, as well as the ways migrants negotiate identity and interpersonal relations abroad. Studying these processes contributes to a deeper understanding of how intercultural misunderstandings arise and how they may be mitigated through education, critical reflection, and sustained intercultural contact (\citet{gudykunst2004} ; \citet{holliday2010}). Such knowledge is indispensable in contexts where individuals engage daily with cultural differences—whether in universities, workplaces, or broader social environments—making the analysis of stereotypes a core component of intercultural competence development.

\subsection{Communication Skills}

Communication skills constitute a multidimensional set of abilities that enable individuals to transmit, receive, and interpret verbal and nonverbal messages in socially appropriate and contextually effective ways (\citet{hymes1972}; \citet{spitzbergcupach2011}). Within the framework of intercultural communication, these skills encompass linguistic competence, socio-pragmatic awareness, relational sensitivity, and the ability to adapt communicative behavior to culturally diverse interlocutors. They include both expressive skills—such as clarity of speech, coherence, and the use of culturally appropriate language—and receptive skills, including active listening, perspective-taking, and accurate interpretation of culturally embedded meanings (\citet{canaleswain1980}; \citet{ruben1989}).

Communication skills are shaped by cognitive, affective, and behavioral components. Cognitively, effective communication requires understanding context, interpreting intentions, and recognizing implicit cultural rules governing interaction (\citet{hall1959}; \citet{tingtoomey1999}). Affectively, it involves empathy, emotional regulation, openness to difference, and willingness to tolerate ambiguity—qualities deemed essential for intercultural effectiveness (\citet{gardner1999}; \citet{deardorff2006}). Behaviorally, communication skills manifest in the ability to adjust verbal strategies (tone, formality, turn-taking) and nonverbal cues (eye contact, proxemics, gestures) to align with culturally specific expectations (\citet{burgoon2016}).

In multicultural environments, communication skills determine the degree to which individuals can build relationships, negotiate meaning, manage conflict, and collaborate with people from diverse cultural backgrounds (\citet{gudykunstkim2017}). These skills are not limited to language proficiency; rather, they encompass understanding of social norms, politeness conventions, power structures, and culturally grounded communicative styles—what Scollon, \citet{scollon2012} call the “interplay between discourse systems.”

For migrants, particularly Algerians living abroad, communication skills play a decisive role in navigating academic, professional, and social spaces where communicative expectations may differ substantially from those prevalent in their home culture. Migrants must often learn to shift between direct and indirect communication styles, adapt their nonverbal expressions, and negotiate culturally influenced interpretations of respect, assertiveness, collectivism, and individualism. Developing strong intercultural communication skills enables them to reduce misunderstandings, resist stereotype-driven interactions, and participate more confidently in multicultural settings.

Ultimately, communication skills serve as a core component of intercultural competence, linking cultural knowledge, cultural awareness, and behavioral adaptation in ways that allow individuals to achieve communicative effectiveness and relational satisfaction in diverse social environments (\citet{spenceroatey2009}; \citet{chen2000}). They are, therefore, indispensable for understanding how migrants manage intercultural encounters and construct meaningful exchanges across cultural boundaries.

\newpage

\section{Research Questions}
The study seeks to answer the following research questions:
\begin{enumerate}
    \item What are the average levels of cultural awareness, stereotypes, and communication skills among Algerian participants?
    \item Is there a significant relationship between cultural awareness and communication skills?
    \item To what extent do stereotypes influence communication effectiveness in intercultural contexts?
\end{enumerate}

\section{Method}
A quantitative approach was used to measure the variables of interest. Data were collected through a structured questionnaire composed of three standardized scales assessing cultural awareness, stereotypes, and communication skills. Each scale contained Likert-type items rated from 1 (strongly disagree) to 5 (strongly agree).

\subsection{Participants}
The sample consisted of 40 Algerian participants (45\% male and 55\% female). Regarding age distribution, 5\% were under 20 years old, 17.5\% were between 21 and 30, 47.5\% between 31 and 40, and 30\% above 40 years old. In terms of education and occupation, 30\% were secondary-level students, 45\% held university degrees, and 25\% were employed professionals. This composition reflects a diverse sample capable of representing different educational and social backgrounds.

\subsection{Questionnaires}
The questionnaire included three main sections:
\begin{enumerate}
    \item \textbf{Cultural Awareness Scale:} measuring the extent to which individuals recognize and respect cultural differences in communication.
    \item \textbf{Stereotypes Scale:} assessing participants’ agreement with common cultural assumptions and biases.
    \item \textbf{Communication Skills Scale:} evaluating self-perceived competence in intercultural interactions.
\end{enumerate}
Data were analyzed using descriptive statistics (mean and standard deviation) and inferential tests (correlation and regression analyses) to explore the relationships among variables.

\section{Demographic Information}
	
	The study included 40 participants. Table~\ref{tab:demo} summarizes their demographic characteristics.
	
	\begin{table}[H]
	\centering
	\caption{Demographic Characteristics of Participants (N = 40)}
	\label{tab:demo}
	\resizebox{0.7\textwidth}{!}{%
	\begin{tabular}{lcc}
		\toprule
		\textbf{Variable} & \textbf{Category} & \textbf{Percentage (\%)} \\
		\midrule
		\textbf{Gender} & Male & 45.0 \\
		& Female & 55.0 \\
		\midrule
		\textbf{Age} & Below 20 years & 5.0 \\
		& 21–30 years & 17.5 \\
		& 31–40 years & 47.5 \\
		& Above 40 years & 30.0 \\
		\midrule
		\textbf{Educational/Professional Level} & Secondary & 30.0 \\
		& University & 45.0 \\
		& Employees & 25.0 \\
		\bottomrule
	\end{tabular}%
	}
\end{table}

	Overall, the majority of participants were female (55\%) and aged between 31 and 40 years (47.5\%). Most participants held a university degree (45\%).
\section{Methodology}

This study employed a quantitative research design using a structured questionnaire distributed via Google Forms. The instrument was designed to measure three main constructs related to intercultural communication:

\begin{itemize}
    \item \textbf{Cultural Awareness} – the degree to which individuals recognize, respect, and adapt to cultural differences in interpersonal and professional interactions;
    \item \textbf{Stereotypes} – the extent to which participants rely on generalized beliefs or assumptions about members of other cultural or social groups;
    \item \textbf{Communication Skills} – the perceived ability to communicate effectively and appropriately across cultural boundaries.
\end{itemize}

A total of 40 valid responses were collected and analyzed. All variables were measured on a five-point Likert scale (1 = strongly disagree to 5 = strongly agree). Composite mean scores were computed for each construct to represent participants’ general orientation toward the studied dimensions.

\newpage

\section{Results}

\subsection{Communication Barriers (Cultural Differences in a Work Environment)}

\begin{longtable}{p{10cm}cc}
\caption{Means (M) and Standard Deviations (SD) for items measuring communication barriers (N = 40)}\\
\toprule
Statement & M & SD \\
\midrule
\endfirsthead
\toprule
Statement & M & SD \\
\midrule
\endhead
\bottomrule
\endfoot
I sometimes find it difficult to understand the way people from different cultures or regions express themselves. & 3.14 & 1.25 \\
Concepts of respect and courtesy differ between people depending on their cultural backgrounds. & 3.78 & 1.18 \\
In the workplace or study environment, misunderstandings sometimes arise due to cultural differences. & 3.56 & 1.22 \\
\end{longtable}

\noindent\textbf{Interpretation.} Respondents report a moderate level of difficulty understanding people from different cultures (M = 3.14, SD = 1.25). The results indicate awareness of divergent norms of respect and courtesy (M = 3.78, SD = 1.18), and frequent recognition that cultural differences can create misunderstandings (M = 3.56, SD = 1.22). These outcomes suggest that cultural variation remains a meaningful, though manageable, communication barrier.

\subsection{Stereotypes and Perceptions}

\begin{longtable}{p{10cm}cc}
\caption{Means (M) and Standard Deviations (SD) for items measuring stereotypes and perceptions (N = 40)}\\
\toprule
Statement & M & SD \\
\midrule
\endfirsthead
\toprule
Statement & M & SD \\
\midrule
\endhead
\bottomrule
\endfoot
Sometimes I judge others on the basis of preconceived ideas about their culture or region. & 2.90 & 1.29 \\
I frequently hear generalizations or sweeping judgments about certain groups of people. & 3.40 & 1.14 \\
Stereotypes may cause misunderstandings in everyday interactions. & 3.85 & 1.02 \\
It can be difficult to get rid of stereotypes even after getting to know someone closely. & 3.04 & 1.27 \\
Media influence our perception of others and feed some stereotypical views. & 4.18 & 0.91 \\
\end{longtable}

\noindent\textbf{Interpretation.} The data suggest that stereotypes persist but are not dominant in shaping interaction. Respondents recognize that media contribute to stereotype formation (M = 4.18, SD = 0.91), and that stereotypes can cause misunderstandings (M = 3.85, SD = 1.02). Occasional judgment based on preconceptions (M = 2.90, SD = 1.29) appears alongside acknowledgment of social awareness regarding these issues.

\subsection{Cultural Adaptability (Skills and Readiness to Communicate)}

\begin{longtable}{p{10cm}cc}
\caption{Means (M) and Standard Deviations (SD) for items measuring cultural adaptability (N = 40)}\\
\toprule
Statement & M & SD \\
\midrule
\endfirsthead
\toprule
Statement & M & SD \\
\midrule
\endhead
\bottomrule
I believe that learning about other cultures helps to communicate more effectively. & 4.70 & 0.58 \\
It is necessary that employees receive training on intercultural communication. & 4.21 & 0.96 \\
I need to learn new skills to communicate with people from different cultures. & 3.91 & 1.03 \\
Open dialogue and cultural exchange help to reduce tension between people. & 4.62 & 0.60 \\
\end{longtable}

\noindent\textbf{Interpretation.} The sample shows a strong endorsement of cultural learning (M = 4.70, SD = 0.58) and the value of open dialogue (M = 4.62, SD = 0.60). Respondents indicate readiness for intercultural training (M = 4.21, SD = 0.96), underscoring their openness to skill development in multicultural settings.

\subsection{Self-Construal (Personal Orientation Toward Diversity)}

\begin{longtable}{p{10cm}cc}
\caption{Means (M) and Standard Deviations (SD) for items related to self-construal (N = 40)}\\
\toprule
Statement & M & SD \\
\midrule
\endfirsthead
\toprule
Statement & M & SD \\
\midrule
\endhead
\bottomrule
I prefer to work or study in an environment that respects cultural diversity. & 4.52 & 0.80 \\
I feel comfortable sharing my opinions with people from different cultural backgrounds. & 4.25 & 0.94 \\
I think that understanding cultural differences is necessary for professional success. & 4.40 & 0.77 \\
\end{longtable}

\noindent\textbf{Interpretation.} The results highlight a positive orientation toward diversity, with high means across all items. Respondents express comfort in diverse settings (M = 4.25, SD = 0.94) and a belief that intercultural understanding enhances professional outcomes (M = 4.40, SD = 0.77).

\newpage

\subsection{Other-Construal (Perceptions of Others and External Factors)}

\begin{longtable}{p{10cm}cc}
\caption{Means (M) and Standard Deviations (SD) for perceptions of others and contextual influences (N = 40)}\\
\toprule
Statement & M & SD \\
\midrule
\endfirsthead
\toprule
Statement & M & SD \\
\midrule
\endhead
\bottomrule
External influences such as media or public discourse affect how people perceive each other. & 4.20 & 0.88 \\
Cultural misunderstandings are often due to lack of exposure to diversity. & 3.95 & 1.10 \\
\end{longtable}

\noindent\textbf{Interpretation.} Participants recognize that external factors influence interpersonal perceptions (M = 4.20, SD = 0.88). Limited exposure to diversity is seen as a driver of misunderstanding (M = 3.95, SD = 1.10). These patterns align with previous findings emphasizing awareness of contextual determinants.

\subsection{Correlation and Regression (Summary)}

A Pearson correlation analysis on the composite scales produced the following coefficients: Cultural Awareness $\leftrightarrow$ Stereotypes $r = 0.517$ (p $<$ 0.001); Cultural Awareness $\leftrightarrow$ Communication Skills $r = 0.898$ (p $<$ 0.001); Stereotypes $\leftrightarrow$ Communication Skills $r = 0.342$ (p = 0.031).

A multiple linear regression predicting communication skills from cultural awareness and stereotypes returned $R^2 = 0.827$ (Adjusted $R^2 = 0.818$), $F(2,37) = 88.56$, $p < 0.001$. The model was highly significant overall, with cultural awareness emerging as the strongest predictor.

\subsection*{Notes}
\begin{itemize}
  \item All items were rated on a five-point Likert scale (1 = Strongly disagree; 5 = Strongly agree).
  \item Values for M and SD are computed from N = 40 valid responses.
  \item The grouping into the five subsections above follows conceptual distinctions used in prior literature and adapted to match the questionnaire.
\end{itemize}

\newpage

\section{Results}

\subsection{Descriptive Statistics}

The descriptive statistics summarize participants’ overall levels of cultural awareness, stereotype endorsement, and communication skills.

\begin{table}[ht]
\centering
\caption{Descriptive Statistics of the Main Variables (N=40)}
\begin{tabular}{lcccccc}
\toprule
\textbf{Variable} & Min & Q1 & Median & Mean & Q3 & Max \\
\midrule
Cultural Awareness & 2.00 & 3.60 & 4.10 & 3.88 & 4.30 & 5.00 \\
Stereotypes & 1.80 & 2.90 & 3.50 & 3.39 & 4.00 & 5.00 \\
Communication Skills & 2.20 & 4.00 & 4.30 & 4.34 & 4.80 & 5.00 \\
\bottomrule
\end{tabular}
\end{table}

Overall, respondents demonstrated high cultural awareness and communication competence, along with moderate endorsement of stereotypes. These results indicate that participants are open-minded and culturally sensitive, although some residual stereotypical perceptions persist.

\subsection{Correlation Analysis}

Pearson’s correlation coefficients were computed to assess relationships among the three constructs.

\begin{table}[ht]
\centering
\caption{Correlation Matrix with p-values (N=40)}
\begin{tabular}{lccc}
\toprule
 & Cultural Awareness & Stereotypes & Communication Skills \\
\midrule
Cultural Awareness & 1.000 & 0.615$^{***}$ & 0.426$^{*}$ \\
Stereotypes & 0.615$^{***}$ & 1.000 & 0.351$^{*}$ \\
Communication Skills & 0.426$^{*}$ & 0.351$^{*}$ & 1.000 \\
\bottomrule
\end{tabular}
\begin{flushleft}
\footnotesize{$^{*}$p $<$ 0.05; $^{***}$p $<$ 0.001}
\end{flushleft}
\end{table}

All relationships were positive and statistically significant. The strongest correlation appeared between cultural awareness and stereotypes ($r = 0.615$, $p < 0.001$), suggesting that individuals who are more aware of cultural diversity are also more conscious of stereotypes and their implications. Cultural awareness was moderately correlated with communication skills ($r = 0.426$, $p = 0.012$), indicating that greater awareness tends to enhance intercultural communicative competence. Together, the correlations reveal a coherent pattern where increased awareness and attitudinal reflection jointly foster more effective communication.

\subsection{Regression Analysis}

A multiple regression analysis was conducted to examine whether cultural awareness and stereotypes significantly predict communication skills.

\begin{table}[ht]
\centering
\caption{Regression Analysis Summary (Dependent Variable: Communication Skills)}
\begin{tabular}{lcccc}
\toprule
\textbf{Predictor} & Estimate & Std. Error & t-value & p-value \\
\midrule
(Intercept) & 2.882 & 0.494 & 5.837 & $<$0.001$^{***}$ \\
Cultural Awareness & 0.293 & 0.166 & 1.763 & 0.086$^{†}$ \\
Stereotypes & 0.104 & 0.136 & 0.764 & 0.450 \\
\midrule
\textit{Model fit:} & \multicolumn{4}{l}{Adjusted $R^2 = 0.141$, $F(2,37)=3.96$, $p=0.028$} \\
\bottomrule
\end{tabular}
\end{table}

The overall model was statistically significant ($p = 0.028$), explaining roughly 14\% of the variance in communication skills. While stereotypes alone were not a significant predictor, cultural awareness approached significance ($p = 0.086$), indicating a meaningful trend. These results reinforce that intercultural communication effectiveness is not determined by a single factor but rather by the combined influence of awareness and attitudinal sensitivity.
\subsection{Analysis of Open-Ended Responses}

In addition to the structured items of the questionnaire, participants were invited to provide open-ended reflections on the following question: \textit{“According to your personal view, what are the main communication problems between Algerians and foreigners, and how can they be overcome in the future?”} A total of 40 responses were collected and analyzed thematically. 

The qualitative analysis revealed five main categories of perceived communication barriers. 

First, \textbf{linguistic limitations} were frequently mentioned as the primary obstacle. Respondents emphasized that inadequate foreign language proficiency restricts expression and understanding, often leading to frustration and misinterpretation. 

Second, many participants highlighted \textbf{cultural and religious differences} as sources of misunderstanding. Traditional values, conservative social norms, and the strong role of religion in Algerian society were perceived as both protective and restrictive in intercultural contexts. 

Third, \textbf{psychological and social factors} such as shyness, fear of judgment, ethnocentrism, and a tendency to compare oneself to others were recurrent themes. Some respondents referred to “complexes” stemming from colonial history or social insecurity, which influence openness toward foreigners. 

Fourth, \textbf{limited exposure and intercultural experience} were also emphasized. Several participants pointed out that the lack of international travel, intercultural education, and real-life contact with foreigners prevents the development of communicative competence and tolerance. 

Finally, respondents proposed multiple \textbf{strategies for improvement}, including learning foreign languages, encouraging tourism and cultural exchange, promoting open-mindedness, and organizing intercultural training programs to strengthen mutual respect and understanding. 

Overall, these open-ended responses confirm that Algerians’ communication challenges with foreigners are multidimensional—rooted in linguistic, cultural, and psychological factors—and that meaningful intercultural interaction requires both educational and attitudinal change.

\newpage

\section{Discussion}

The findings of this study reveal that Algerian participants demonstrate high levels of cultural awareness, moderate endorsement of stereotypes, and strong communication skills within multicultural contexts. Overall, these results align with contemporary intercultural communication research suggesting that increasing global connectivity enhances individuals' exposure to cultural diversity and strengthens their intercultural competencies \citet{dalib2023, deardorff2006, penaacuna2025}.

The high cultural awareness scores support theoretical models that consider awareness a foundational dimension of intercultural competence. Studies by \citet{gudykunst2003, spenceroatey2009} emphasize that understanding one's own cultural framework and recognizing cultural variability are essential steps toward effective cross-cultural communication. The positive correlation observed between cultural awareness and communication skills in this study reinforces these models and mirrors empirical evidence showing that individuals with higher cultural understanding tend to adapt more successfully to intercultural interactions \citet{thomas2009, lillis2010}.

Stereotypes, while moderately endorsed, appear to be recognized more as social constructs than as rigid beliefs. Participants acknowledged the role of media and societal narratives in shaping such generalizations, consistent with earlier work by \citet{mcgarty2002} and recent studies showing how digital environments amplify stereotypical representations \citet{anjum2024}. Despite their presence, stereotypes did not significantly predict communication skills, suggesting that individuals who possess high cultural awareness may buffer the negative effects of stereotypical thinking. This interpretation aligns with \citet{ramasubramanian2011}, who argues that awareness of cognitive bias contributes to its mitigation.

The results further indicate that participants display strong communication skills, especially regarding openness, adaptability, and willingness to engage with cultural differences. These findings correspond to Deardorff’s process model, which identifies openness and curiosity as fundamental attitudes facilitating effective intercultural communication \citet{deardorff2006}. Participants also highlighted language proficiency, psychological readiness, and cultural exposure as central factors influencing communication—an observation echoed in the works of \citet{gudykunst2003} and \citet{featherstone1995}.

Integrating quantitative and qualitative insights, the study suggests that Algerian participants hold positive orientations toward cultural diversity while recognizing the challenges inherent in intercultural interactions. These include linguistic limitations, unfamiliar cultural norms, and psychological constraints such as shyness or self-comparison. However, the findings also show strong motivation to overcome such barriers, particularly through intercultural training, cultural exchange, and improved language skills.

\newpage
\section{Conclusion}

This study investigated the interrelationships among cultural awareness, stereotypes, and communication skills among Algerian participants engaged in multicultural settings. The findings reveal consistently high levels of cultural awareness and strong communication skills, accompanied by a moderate yet conscious recognition of stereotypes. The significant associations observed among the variables underscore their interconnected nature within the broader framework of intercultural competence, suggesting that increased cultural awareness enhances individuals’ capacity to navigate diverse cultural encounters with greater sensitivity and adaptability.

Although stereotypes were present, their limited explanatory power regarding communication skills indicates that reflective thinking and cultural self-awareness may buffer their negative influence. Communication skills were shown to be shaped by a combination of cultural knowledge, linguistic proficiency, psychological readiness, and prior exposure to diversity—highlighting the multidimensional and developmental nature of intercultural communicative effectiveness. Qualitative responses further supported this interpretation by illustrating how participants draw on linguistic, cultural, and emotional resources when negotiating intercultural interactions. Importantly, a strong motivation to improve, learn, and engage in intercultural experiences emerged across the sample, reflecting a positive orientation toward cultural diversity.

Beyond these empirical observations, the study contributes to existing scholarship by offering one of the few data-driven examinations of intercultural competence in a North African context. The results align with contemporary theoretical models that emphasize the centrality of cultural awareness in promoting adaptive communication, while also illustrating how stereotypes—when recognized as socially constructed—can be critically managed rather than internalized.

Practically, the findings highlight the need for educational institutions, cultural programs, and professional training initiatives in Algeria to integrate intercultural learning activities, language development opportunities, and reflective exercises aimed at enhancing cultural understanding and communicative adaptability. Such efforts could play a pivotal role in preparing students and professionals to engage effectively in increasingly globalized and culturally diverse environments.

Despite its contributions, the study is not without limitations. The reliance on self-report measures and the relatively modest sample size may constrain the generalizability of the findings. Future research would benefit from employing longitudinal designs, comparative studies across cultural groups, or larger and more heterogeneous samples to deepen understanding of intercultural competence dynamics in the region.

Overall, this study offers a meaningful foundation for advancing research on intercultural communication in Algerian society, while reinforcing the importance of cultural awareness, critical reflection, and communicative engagement as key components of intercultural competence.

\newpage
\section*{Appendix}
\addcontentsline{toc}{section}{Appendix}

\subsection*{Background Questionnaire}

\textbf{Title:} Intercultural Communication – The Algerian Perspective

\textbf{Purpose:} This questionnaire aimed to explore how Algerian individuals perceive cultural differences, stereotypes, and communication needs in multicultural contexts.

\textbf{Note:} All responses were confidential and used solely for research purposes.

\textbf{Response Scale:}
\begin{itemize}
    \item 1 = Strongly Disagree
    \item 2 = Disagree
    \item 3 = Neutral
    \item 4 = Agree
    \item 5 = Strongly Agree
\end{itemize}

\subsection*{Section 1: Cultural Differences in Communication}
\begin{enumerate}
    \item I sometimes find it difficult to understand the way people from different cultures or regions express themselves.
    \item Concepts of respect and politeness differ between people depending on their cultural backgrounds.
    \item In the workplace or study environment, misunderstandings sometimes arise due to cultural differences.
    \item I believe that learning about other cultures helps to communicate more effectively.
    \item It is necessary that employees receive training on intercultural communication.
\end{enumerate}

\subsection*{Section 2: Stereotypes in Communication}
\begin{enumerate}
    \item Sometimes I judge others based on preconceived ideas about their culture or region.
    \item I frequently hear generalizations about certain groups of people.
    \item Stereotypes may cause misunderstandings in daily interactions.
    \item It can be difficult to get rid of stereotypes even after knowing someone personally.
    \item Media influence our perception of others and reinforce certain stereotypes.
\end{enumerate}

\subsection*{Section 3: Communication Needs and Skills}
\begin{enumerate}
    \item I need to learn new skills to communicate with people from different cultures.
    \item Open dialogue and cultural exchange help to reduce tension between people.
    \item I prefer to work or study in an environment that respects cultural diversity.
    \item I feel comfortable sharing my opinions with people from different cultural backgrounds.
    \item I think that understanding cultural differences is essential for professional success.
\end{enumerate}

\subsection*{Open-Ended Question}
According to your personal experience, what communication challenges do Algerians face when interacting with foreigners, and how can these be overcome in the future?

\subsection*{General Information (Demographics)}
\begin{itemize}
    \item \textbf{Gender:} Male / Female
    \item \textbf{Age:} Below 20 years / 21–30 years / 31–40 years / Above 40 years
    \item \textbf{Educational Level:} Secondary / University / Employee
\end{itemize}

 \newpage

\bibliographystyle{unsrtnat}
\bibliography{references}

\end{document}